\documentclass[
jctc,
preprint,
numerical,
floatfix
]{revtex4-1}

\usepackage{graphicx}
\usepackage{xcolor}
\usepackage{amsmath}
\usepackage[colorlinks=true, linkcolor=green, citecolor=green]{hyperref}

\renewcommand{\selectlanguage}[1]{}

\makeatletter
\def\fps@figure{t}
\AtBeginDocument{\let\LS@rot\@undefined}
\makeatother

\begin{document}

\title{Distilling first-principles accuracy into compact machine learning potentials for condensed-phase chemistry}

\author{Sijia Chen}
\email{schen@flatironinstitute.org}
\affiliation{%
    Initiative for Computational Catalysis, Flatiron Institute, New York, New York 10010, United States
}

\author{Niamh O’Neill}
\affiliation{Max Planck Institute for Polymer Research, Mainz 55128, Germany}

\author{Benjamin X. Shi}
\affiliation{%
    Initiative for Computational Catalysis, Flatiron Institute, New York, New York 10010, United States
}

\author{Venkat Kapil}%
\email{v.kapil@ucl.ac.uk}

\affiliation{%
    Department of Physics and Astronomy, University College London, 7-19 Gordon St, London WC1H 0AH, UK
}
\affiliation{%
   Thomas Young Centre and London Centre for Nanotechnology, 
   9 Gordon St, London WC1H 0AH, UK
}

\begin{abstract}

Accurate machine learning interatomic potentials (MLIPs) have made first-principles-quality potential energy surfaces increasingly accessible for condensed-phase chemistry, but their inference cost can still limit the sampling needed to compute experimentally relevant observables.
In this work, we combine transfer learning and knowledge distillation to construct compact ``student" models that retain the accuracy of much larger ``teacher" models obtained by applying transfer learning to foundation models.
The resulting students reduce production simulation cost by approximately an order of magnitude, making high-accuracy sampling practical for challenging condensed-phase problems.
We demonstrate this across three problems of increasing sampling complexity: finite-temperature $NPT$ simulations of ice Ih, classical and path-integral simulations of liquid water over 240--370~K, and path-integral umbrella-sampling simulations of water dissociation at the anatase TiO$_2$(101)/water interface.
In all cases, the distilled students reproduce the target observables of their teachers more reliably than models of the same size trained directly on the limited reference data.
The liquid-water student, distilled from a $\Delta$-learned CCSD(T)-quality teacher, reproduces thermodynamic, structural, transport, and nuclear quantum properties over the full temperature range studied.
At the TiO$_2$/water interface, distillation makes PIMD umbrella sampling practical and shows that nuclear quantum effects lower the dissociation barrier by approximately 2~kcal/mol and shift the molecular--dissociated free energy difference into quantitative agreement with recent solid-state $^{17}$O NMR measurements.
Our work demonstrates how knowledge distillation can make accurate MLIPs practical for the sampling methods needed to connect condensed-phase reaction thermodynamics with experiment, notably for interfacial chemistry and catalysis.
\end{abstract}
	
\maketitle
	
\newpage

\section{Introduction}

\noindent Machine learning interatomic potentials (MLIPs)~\cite{deringer_gaussian_2021, behlerFourGenerationsHighDimensional2021a} have become a cornerstone of chemical and materials science. 
However, despite greatly reducing the computational cost of sampling first-principles quality potential energy surfaces, many problems of interest are still computationally intractable. 
This limitation arises because the model must simultaneously satisfy two stringent requirements.
The machine learning architecture must be expressive enough to accurately describe a wide region of configuration space relevant to a problem but also efficient enough to sample this space in a feasible amount of computational time.\cite{zuoPerformanceCostAssessment2020,fuForcesAreNot2023,leimerothMachinelearningInteratomicPotentials2025,winesCHIPSFFEvaluatingUniversal2025}
Due to the general inverse relation between expressivity and efficiency, a number of challenging regimes relevant to chemical and materials simulations arise, including long molecular dynamics (MD) trajectories~\cite{shawAtomicLevelCharacterizationStructural2010}, enhanced sampling of rare events such as bond breaking~\cite{heninEnhancedSamplingMethods2022}, and path integral (PI) MD simulations that account for nuclear quantum effects (NQEs)~\cite{marklandNuclearQuantumEffects2018}. \\

\noindent In the past few years, there has been significant progress on improving the accuracy of MLIPs across configurational and alchemical space~\cite{jacobs_practical_2025}.
Besides the development of new graph neural network architectures~\cite{batznerE3equivariantGraphNeural2022a, Batatia2022mace}, transfer learning~\cite{thrun_is_1995} -- adapting or exploiting the weights of a pretrained model for data associated with a new problem -- has proven advantageous over training a new model from scratch on the same data.
This approach has enabled MLIPs with beyond-DFT accuracy developed by finetuning a model trained on DFT to small volumes of beyond-DFT data~\cite{smithApproachingCoupledCluster2019a,chenDataEfficientMachineLearning2023} or by augmenting a model trained on DFT with a differential learning~\cite{ramakrishnanBigDataMeets2015} model trained on the difference between DFT and beyond-DFT~\cite{nandiDmachineLearningPotential2021,daru_coupled_2022,oneillRoutineCondensedPhase2025,qu_gold-standard_2025}. 
More recently, transfer learning has simplified the development of accurate MLIPs at (beyond-)DFT accuracy for generic systems by applying various types of finetuning~\cite{kaurDataefficientFinetuningFoundational2025,radova_fine-tuning_2025,novelliFastFourierFeatures2025b,gawkowski_good_2025, liuFinetuningUniversalMachinelearned2026}, or in some cases differential learning~\cite{christiansenDmodelCorrectionFoundation2025,pitfieldActiveDlearningUniversal2026}, to large models trained on diverse datasets, referred to as foundation MLIPs~\cite{chanussotOpenCatalyst20202021, chenUniversalGraphDeep2022, dengCHGNetPretrainedUniversal2023, parkScalableParallelAlgorithm2024, batatiaFoundationModelAtomistic2025a, yuanFoundationModelsAtomistic2026, woodUMAFamilyUniversal2026}.
While foundation MLIPs expedite model development by reducing human effort and training-data generation, their inference cost is typically much higher than that of a carefully curated small MLIP with a simpler architecture trained on a large amount of data~\cite{chengEvidenceSupercriticalBehaviour2020,kapilFirstprinciplesPhaseDiagram2022,qamarAtomicClusterExpansion2023,willmanAtomicClusterExpansion2024,songGeneralpurposeMachinelearnedPotential2024,zhouMillionatomHeatTransport2025}.
As a result, their application to large systems (thousands to millions of atoms), long simulations (nanoseconds to milliseconds~\cite{shawAtomicLevelCharacterizationStructural2010}), or sophisticated sampling algorithms remains challenging.\\

\noindent More recently, efforts have intensified towards reducing the computational cost of MLIPs, ideally at constant accuracy.
An attractive approach, amongst others, is knowledge distillation, a strategy widely used in computer science~\cite{buciluaModelCompression2006,hintonDistillingKnowledgeNeural2015,gouKnowledgeDistillationSurvey2021}.
Within this approach, a large and/or broadly applicable so-called ``teacher'' model is used to generate training data relevant to the problem of interest at the cost of MLIP inference rather than additional first-principles calculations, and a compact ``student'' model with a low inference cost is then trained to fit the problem-specific data. 
In the context of condensed-phase simulations, knowledge distillation has been successfully demonstrated to reduce the inference cost of foundation MLIPs for specific classes of systems~\cite{aminFastSpecializedMachine2025a,cattinAcceleratingMolecularDynamics2026a,gouraudFasterMolecularDynamics2026} and has also been explored to accelerate molecular dynamics driven by fine-tuned foundation MLIPs for materials~\cite{matsumuraKnowledgeDistillationFramework2025b,gardnerDistillationAtomisticFoundation2025a}. \\

\noindent At the same time, there are problems for which the application of knowledge distillation is non-trivial, because it is difficult to generate an exhaustive dataset \textit{a priori}. 
The student model must then be accurate not only on the distillation set, but also for configurations encountered during production simulations, while still retaining low inference cost to maintain the computational advantage of knowledge distillation.
Some examples include chemical reactions in the bulk or at interfaces where a failure of knowledge distillation has been noted~\cite{gardnerDistillationAtomisticFoundation2025a}.
In this case, knowledge distillation would in principle require thorough sampling of transition-state regions in the training set, but generating such data with the teacher model can itself be computationally prohibitive. 
Another is the incorporation of NQEs, which requires expensive PIMD sampling of regions of configuration space (due to zero-point motion and tunnelling) that are much less or not accessible to classical simulations.
In such cases, the challenge is not only how to distill a compact and efficient student, but also how to obtain an accurate teacher in a data-efficient manner, since the accuracy ultimately achievable by the student is inherited from the teacher.
Thus, an understanding of how accurate the teacher must first be, how much of the relevant configuration space the student must see during training, and how expressive the student model itself must be, would be valuable for enabling fast and accurate MLIPs for challenging condensed-phase chemistry. \\
%

\noindent In this work, we combine transfer learning and knowledge distillation to show that compact MLIPs can be made accurate and computationally efficient, while remaining applicable beyond the regimes explicitly represented in the training set.
%
%
We demonstrate this on three condensed-phase problems of progressively increasing difficulty: (1) ice~Ih, where a fine-tuned teacher is distilled into a student that is nearly 10 times faster while accurately predicting the $NPT$ density at thermodynamic state points outside the training regime; (2) liquid water, where a student distilled from a larger $\Delta$-learned CCSD(T)-accurate teacher is fast enough to explore the temperature dependence of the density isobar, diffusion coefficient and quantum kinetic energies over 240-370\,K while incorporating quantum nuclear effects (each requiring large-scale, computationally intensive simulations to converge accurately) in excellent agreement with the experiments;  and (3) proton transfer at the anatase TiO$_2$(101)/water interface, where a student distilled from a DFT fine-tuned teacher is about 10 times faster and enables us to examine how quantum nuclear motion modifies the reaction free energy landscape in qualitative agreement with an experiment.
In each case, we analyse, where relevant, how the teacher should be constructed through transfer learning, how the distillation training set should be sampled, and how the student model should be chosen to provide the most useful balance between expressivity and simulation throughput for a problem at hand.
We find that these trade-offs become progressively tighter as the target problem moves from predicting bulk thermodynamic properties to problems requiring enhanced sampling and/or quantum nuclear motion, but that they nevertheless remain navigable.
Overall, our results show that knowledge distillation can make many condensed-phase chemistry problems tractable that would otherwise be limited by the cost of fine-tuned or $\Delta$-learned models.

\section{Results and Discussion}\label{sec:rad}

\noindent We consider three condensed-phase systems of progressively increasing difficulty: a crystalline solid in the $NPT$ ensemble, a molecular liquid with quantum nuclear effects, and a reactive solid--liquid interface with quantum nuclear effects.
While we briefly summarize the methodology and focus on the key results below, we refer the reader to the Methods section for the associated computational details. 
%


\subsection{Predicting the density of ice Ih}

\begin{figure}[h]
\centering
\includegraphics[width=0.85\textwidth]{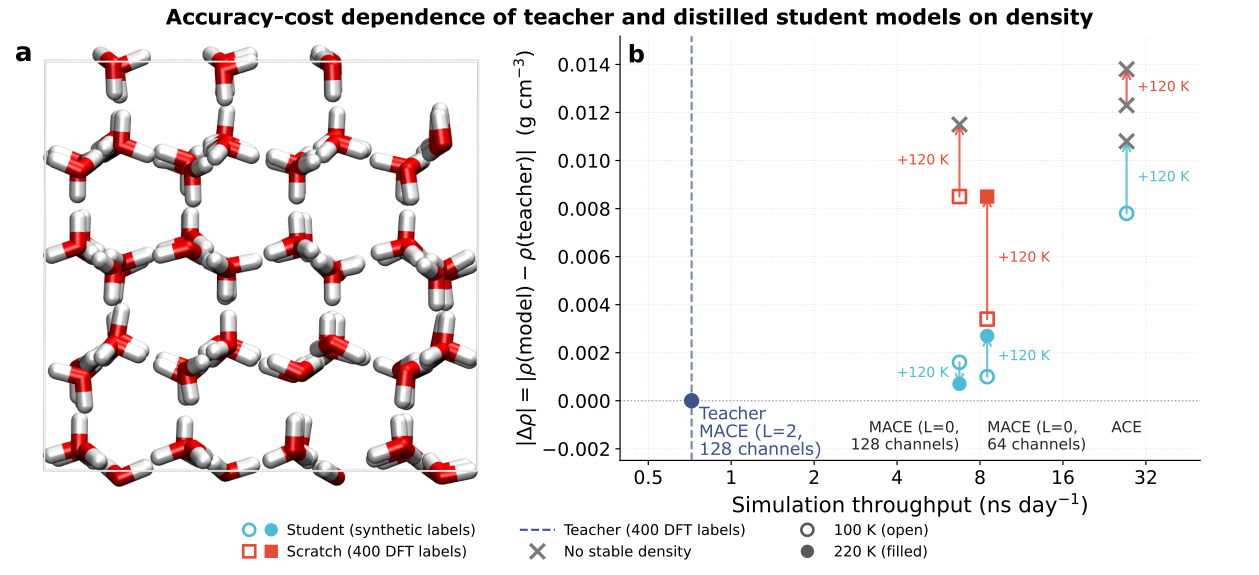}
\caption{Speed--accuracy tradeoff for ice Ih. (a) Representative atomistic configuration of the 384-atom ice Ih supercell used for the NPT simulations, with oxygen atoms in red and hydrogen atoms in white. (b) Absolute deviation of the predicted density from the finetuned teacher, $|\Delta\rho| = |\rho_\mathrm{model} - \rho_\mathrm{teacher}|$, versus simulation throughput (ns~day$^{-1}$, log$_2$ scale). The teacher (dark blue dashed reference line and dot; labelled ``MACE, L=2, 128 channels'') is the finetuned MACE-MP-0a(L) foundation model and defines $|\Delta\rho|=0$ at its measured throughput. The three small student architectures (labelled ``MACE, L=0, 128 channels'', ``MACE, L=0, 64 channels'', and ``ACE'') are trained on either 2000 teacher-labelled configurations (Student, cyan circles) or the original 400 DFT configurations (Scratch, orange-red squares). Open markers correspond to 100~K, 1~bar; filled markers to 220~K, 1~bar. Vertical segments connect the two temperatures of the same model. Configurations that fail to sustain stable NPT dynamics with the default 0.5~fs timestep are marked as ``No stable density'' (gray X) and stacked above the architecture's stable points. All simulations use a 384-atom supercell on a single NVIDIA A100-PCIe GPU; replica-to-replica standard deviations of the density are below $0.0003\ \mathrm{g\ cm^{-3}}$ for all stable runs and are smaller than the symbol size.}
\label{fig:ice_density}
\end{figure}

\noindent We first demonstrate the combination of transfer learning and knowledge distillation on a simple solid but for a challenging observable, both within and outside the regime represented in the training set.
We consider the  ambient-pressure density of ice Ih, a prototypical system which has served as a test bed for general molecular systems.~\cite{kaurDataefficientFinetuningFoundational2025}
Methods capable of resolving the finite-temperature properties of ice Ih in the constant-pressure ensemble are applicable to a much broader class of organic molecular crystals~\cite{pia_accurate_2025}. 
Moreover, density is a particularly challenging observable to converge with MLIPs, and small energy and force errors alone do not guarantee its accurate prediction.~\cite{kaurDataefficientFinetuningFoundational2025,magdau_machine_2023}. \\

\noindent We consider the dataset of ~\citet{kaurDataefficientFinetuningFoundational2025} which corresponds to 500 configurations sampled randomly from (coarsely-converged DFT settings) first-principles molecular dynamics simulations at 100\,K and 1\,bar. 
We first obtain an accurate teacher model by finetuning the foundation model MACE-MP-0(L) on 400 configurations (from the said dataset) labelled with revPBE-D3(0) total energies and forces, achieving an energy RMSE below 0.01 meV/atom and a force RMSE of 1.04 meV/\AA{} on a test set comprising 100 configurations.
Fig. S1-S4, show the parity/violin plot of energy and force predictions/errors on the test set and demonstrate high fidelity across the full dataset. 
We test this model on the density at 100\,K and 1\,bar by performing NPT simulations. 
We obtain a density of $0.9223 \pm 0.0002\ \mathrm{g\ cm^{-3}}$, in quantitative agreement with the MLIP and DFT reference densities reported in ~\citet{kaurDataefficientFinetuningFoundational2025}, thereby demonstrating the capability of transfer learning in developing an accurate model.
However, this model is relatively large and computational demanding (with 780,698 parameters) owing to its use of $L = 2$ spherical harmonics and 128 channels, and achieves only 0.72 ns/day of MD throughput on our 384-atom ice Ih supercell (Figure~\ref{fig:ice_density}c). 
Since a model of this capacity is intended to describe systems across the full periodic table, it is natural to distil it into a more compact model specialized to ice Ih at ambient pressure. \\

\noindent To generate the training set for knowledge distillation, we consider the original dataset with 400 configurations and add 1600 new configurations by rattling the former as was proposed by ~\citet{gardnerDistillationAtomisticFoundation2025a}. 
The reasoning behind rattling (instead of performing a simulation with the teacher model) is that ice Ih is solid and its thermal fluctuations at low temperatures would mimic a Gaussian distribution. 
We explore three different student architectures in increasing order of expressivity and consequently reducing order of computational efficiency: (1) an ACE model with a Finnis--Sinclair two-density embedding, up to 1000 basis functions per element, and body orders up to 6, comprising 5,296 parameters; (2) a MACE model with invariant ($L = 0$) messages and 64 channels, comprising 148,944 parameters; and (3) a MACE model with invariant ($L = 0$) messages and 128 channels, comprising 460,432 parameters.
Energy and force errors on a held-out test set for all student and scratch architectures are reported in SI Figs.~S1--S4. 
\\

\noindent In Fig.~\ref{fig:ice_density}, we report the error in the density at 100\,K and 1\,bar relative to the teacher model for the three student architectures. 
As a point of comparison, we also include the corresponding results for models of the same architectures trained from scratch on the original 400-configuration dataset, together with their computational cost.
For the least expressive but most computationally efficient ACE models, the model trained from scratch is unstable in production MD, whereas the distilled student, although stabilized by training on a larger volume of data, yields the largest density error among the student models.
By contrast, the more expressive MACE student models, although approximately four times slower than the ACE student, recover the density to within 0.002 $\mathrm{g\ cm^{-3}}$.
The corresponding MACE models trained from scratch, however, exhibit larger errors, with the more expressive 128-channel model again performing worse.
Although these results already show that, at 100\,K, the student models outperform equally sized models trained from scratch, this analysis does not fully reveal the benefits of finetuning and distillation. 
We therefore also evaluate the density error at 220\,K, in a region of configuration space not seen during student training, relative to the teacher.
Under these conditions, both ACE models, whether distilled or trained from scratch, are numerically unstable.
Among the MACE models trained from scratch, the 64-channel model exhibits an approximately twofold increase in the density error at 220~K, whereas the 128-channel model becomes unstable.
A plausible explanation is that the 128-channel scratch model's larger parameter count is underdetermined by only 400 training configurations, leaving rough PES features off-distribution that destabilize NPT dynamics at 220 K.
By contrast, the distilled student models retain errors of the same order of magnitude as at 100 K, indicating robust generalizability beyond the regime represented in the training set. \\

\noindent This system yields two main lessons. First, distillation can produce compact models that are both efficient and accurate, while retaining sufficient generalizability to remain reliable slightly outside the regime represented in the training set. 
In the case of solids, this may be achieved by expanding the reference training set efficiently through rattling and relabeling. 
Second, the most useful student is not necessarily the cheapest or least expressive one. 
Although a very inexpensive model may be sufficient when no exploration beyond the training set is required, in the present case the 64-channel MACE student provides the best balance between accuracy, generalizability, and cost, reproducing the teacher's density predictions at both 100~K and 220~K while remaining approximately 10 times faster than the teacher. 


\subsection{Quantum nuclear effects in ambient pressure water}

\begin{figure}[h]
\centering
\includegraphics[width=1.0\textwidth]{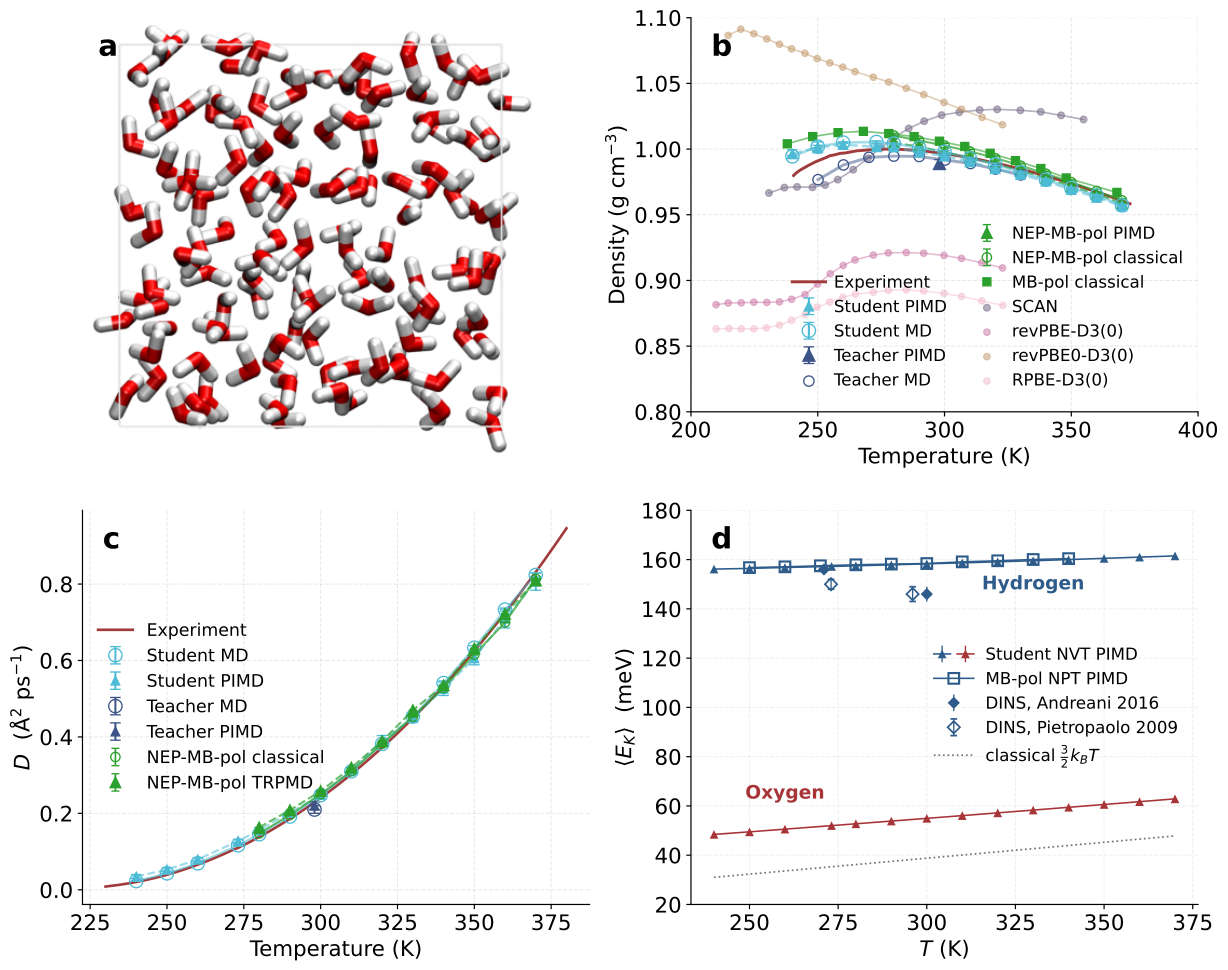}
\caption{Thermodynamic, dynamical, and quantum nuclear properties of liquid water predicted by the student model, compared with experiment and other simulations. (a) Representative atomistic snapshot of a liquid water configuration from the production simulations, with oxygen atoms in red and hydrogen atoms in white. (b) Density isobar. Classical molecular dynamics (MD) and path-integral MD (PIMD) results for the student and teacher are compared with several classical-nuclei DFT references\cite{monterodehijesDensityIsobarWater2024} and experiment.\cite{holtenThermodynamicsSupercooledWater2012a,hareDensitySupercooledWater1987,lideCRCHandbookChemistry2004b} Teacher data are reproduced from O'Neill et al.\cite{oneillRoutineCondensedPhase2025} All D3(0) curves correspond to the zero-damping variant of the Grimme dispersion correction.\cite{grimmeConsistentAccurateInitio2010} (c) Self-diffusion coefficient as a function of temperature, corrected for finite-size effects.\cite{yehSystemSizeDependenceDiffusion2004} Experimental data are from Refs.~\cite{holzTemperaturedependentSelfdiffusionCoefficients2000,priceSelfDiffusionSupercooledWater1999}. (d) Mean nuclear quantum kinetic energy $\langle E_K \rangle$ of hydrogen and oxygen atoms from NVT PIMD simulations (64 beads), compared with MB-pol NPT PIMD results from Kapil et al.\cite{kapilAnisotropyProtonMomentum2018} and deep inelastic neutron scattering (DINS) measurements.\cite{andreaniDirectMeasurementsQuantum2016,pietropaoloQuantumEffectsWater2009} The classical equipartition limit $\frac{3}{2}k_BT$ is shown as a dashed line for each species. MD results are plotted with circles and PIMD results with triangles. Lines between symbols are guides to the eye.}
\label{fig:water_properties}
\end{figure}

\noindent We next simulate static and dynamical properties of liquid water at 1\,bar while explicitly incorporating quantum nuclear effects. 
This is a more challenging case for the combined transfer-learning and distillation strategy for two reasons. 
First, relative to a solid, the liquid spans a broader, more flexible, and more temperature-dependent configuration space, which is further expanded by quantum delocalization of protons~\cite{ceriotti_nuclear_2013}. 
This requires a distilled student that is both sufficiently expressive and efficient. 
Second, the properties of liquid water are highly sensitive to the underlying potential energy surface, including the choice of DFT functional and dispersion correction~\cite{gillanPerspectiveHowGood2016a}. 
This in turn requires a correspondingly accurate teacher. 
To address the said challenges, we consider a CCSD(T) accurate teacher model, trained as a short-ranged differential learning MACE model which augments a DFT-trained baseline MACE model by~\citet{oneillRoutineCondensedPhase2025}.  
The original work explored the classical structural properties of water over the range 250 to 330\,K. 
Classical transport properties and the impact of quantum nuclear effects on structural and dynamical properties were considered only at 298\,K due to the high combined cost of the baseline and learned differential potentials and the 64 times higher computational cost of PIMD than standard MD.
We demonstrate how this model can be distilled into a much faster MLIP suitable for studying the temperature dependence of structural and dynamical properties while including NQEs. \\

\noindent We construct the distillation dataset from equilibrium NVT simulations with the teacher model at 260, 300, 330, and 360 K and densities of 0.85, 0.90, 0.95, 1.00, 1.05, 1.10, and 1.15 $\mathrm{g/cm^3}$, and from equilibrium NPT simulations at the same temperatures and pressures of -1500, -500, -300, 1, 2, 500, 1000, 4000, and 8000 bar. 
Each simulation is run for 0.5 ns, provided that it remains stable. 
To improve coverage of configurations relevant to NQEs, we additionally include 50 ps PIMD NVT simulations at 300 K over densities of 0.85, 0.90, 0.95, 1.00, 1.05, and 1.10 $\mathrm{g/cm^3}$. 
From these trajectories, we select 10,000 teacher-labeled configurations to train a student ACE model using a random 9:1 training:validation split.
This ACE model is parametrized with a Finnis--Sinclair six-density embedding and up to 1000 basis functions per element, and yields a simulation throughput speed-up of 10 compared to the teacher model.
On the validation set, the student model achieves an energy MAE and RMSE of 0.79 and 0.82 meV/atom, respectively, and a force MAE and RMSE of 13.35 and 19.04 meV/\AA, respectively, with respect to the teacher model. 
The validation-set errors are nearly identical to the training-set errors indicating faithful reproduction of the teacher without signs of overfitting.
%
%
As a primary validation, we compare the O--O and O--H radial distribution function at 300\,K obtained with MD and PIMD for the student and teacher models as shown in Fig.~S7 of the SI.
We note the student reproduces the teacher RDFs closely for both classical and quantum nuclei. 
The agreement is especially close for the quantum results, where the student and teacher curves are nearly indistinguishable. 
For the classical RDFs, the student also closely follows the teacher, with the most noticeable difference being a slightly lower first peak in the student O--O RDF. \\

\noindent Having established confidence in the structure of water at ambient conditions, we explore the density of water in the 240--370~K range at 1\,bar. 
To this extent, we perform MD and PIMD simulations in the $NPT$ ensemble. 
At each temperature, five independent MD replicas are run until the density standard deviation falls below 0.001~$\mathrm{g/cm^3}$, requiring trajectory lengths between 2~ns at high temperature and 20~ns at 240~K. 
For PIMD, we perform five independent $NPT$ simulations at each temperature using 32 beads, with trajectory lengths of up to 2~ns per replica.
In Fig.~\ref{fig:water_properties}, we compare the MD and PIMD properties of liquid water predicted by the distilled student with the experiment and the density predicted by the teacher model, taken from Ref.~\citenum{oneillRoutineCondensedPhase2025}, which includes MD simulations from 250-370\,K and only one PIMD simulation at 298\,K. 
For the density isobar (Fig.~\ref{fig:water_properties}b), the student closely reproduces the teacher references at 298\,K to within statistical error.
We also note that the student model reproduced the teacher (up to statistical errors) at temperatures above 280\,K but observe a systematic overestimation of the density as the temperature reduces by up to 0.025~$\mathrm{g/cm^3}$ at 250\,K. 
We attribute this low-temperature discrepancy primarily to the sparse representation of the low-temperature regime in the training set, which contains teacher configurations only down to 260 K and therefore provides limited coverage of low temperature configurations. The combination of limited model expressivity and the limited amount of representative low-temperature configurations leads to the observed discrepancy.
Nonetheless, the student does not deviate from experiment more than the teacher does and also exhibits an improvement over the predictions of MLIPs trained on DFT.
The reduced computational cost of the student further enables extension of PIMD simulations across the full 240-370~K range.
The inclusion of NQEs leads to a small reduction in density, on the order of $\sim 0.004~\mathrm{g/cm^3}$ near room temperature and of similar magnitude across the temperature range studied.
Numerical density values from the classical NPT and PIMD NPT isobars are tabulated in SI Tables~S1 and S2.
%
This subtle difference is noteworthy, as it is consistent with the experimentally observed difference between the densities of H$_2$O and D$_2$O after rescaling D$_2$O by the mass ratio $M_{\mathrm{H_2O}}/M_{\mathrm{D_2O}}$ (SI Table~S3).
Across the 240--370~K range studied, the student's density accuracy is comparable to that of MB-pol~\cite{palosCurrentStatusMBpol2024} and its neuroevolution surrogate NEP-MB-pol~\cite{xuNEPMBpolUnifiedMachinelearned2025}, two of the most accurate molecular models of water currently available.
We next study the self-diffusion coefficient of water with finite-size corrections which was only simulated at 298\,K with the teacher model by ~\citet{oneillRoutineCondensedPhase2025}.
Exploiting the computational efficiency of the student model, we are able estimate the self-diffusion coefficient with MD and PIMD across the full 240 - 370\,K range (Fig.~\ref{fig:water_properties}c). 
At 300~K, the predicted diffusion coefficient agrees with both the teacher and experiment and across 240--370~K, the student remains in close agreement with the experimental temperature dependence, with deviations below 5\%. 
Although experimental self-diffusion coefficients are available for D$_2$O, rescaling them to remove the trivial mass dependence and obtain an H$_2$O-equivalent reference is non-trivial. 
We therefore do not make a direct comparison with experimental D$_2$O diffusion data. 
Numerical diffusion coefficients from the classical NVT and PIMD NVT trajectories are tabulated in SI Tables~S4 and S5. \\
\noindent Across the temperature range studied, the classical and PIMD diffusion coefficients agree within statistical uncertainty, indicating that NQEs have a small effect on water's self-diffusion.
This is consistent with NEP-MB-pol classical and TRPMD simulations~\cite{xuNEPMBpolUnifiedMachinelearned2025}, which similarly show only a small change in the self-diffusion coefficient upon inclusion of NQEs.
By contrast, \textit{ab initio} path-integral simulations of liquid water have reported NQE-induced reductions of the self-diffusion coefficient of approximately 30\% for the revPBE-D3 GGA functional and 14\% for the revPBE0-D3 hybrid functional~\cite{marsalekQuantumDynamicsSpectroscopy2017d}, illustrating that the apparent magnitude of NQEs on water dynamics depends sensitively on the underlying potential energy surface. \\

\noindent We also consider an observable not previously reported for the teacher model and directly sensitive to quantum nuclear motion: the mean nuclear kinetic energy of O and H atoms, $\langle E_K \rangle$ (Fig.~\ref{fig:water_properties}d). 
In contrast to density and diffusion, which are quantitatively affected by NQEs, $\langle E_K \rangle$ differs qualitatively from its classical counterpart. 
For hydrogen, the student predicts $\langle E_K \rangle = 157.2$~meV at 273~K and 158.3~meV at 300~K, exceeding the classical equipartition value by more than a factor of four due to zero-point motion in the high-frequency intramolecular and librational modes. 
The corresponding oxygen kinetic energy is much closer to the classical limit, consistent with its higher mass. 
These values are in quantitative agreement with MB-pol PIMD benchmarks~\cite{chengNuclearQuantumEffects2016, kapilAnisotropyProtonMomentum2018} and with deep inelastic neutron scattering (DINS) measurements~\cite{andreaniDirectMeasurementsQuantum2016} in the supercooled regime, which report $\langle E_K \rangle_\mathrm{H} = 156 \pm 2$~meV at 271~K. 
At room temperature, however, DINS experiments yield a lower value of $146 \pm 3$~meV, highlighting a known discrepancy between experiment and PIMD simulations that remains unresolved~\cite{kapilAnisotropyProtonMomentum2018}. 
The close agreement with both MB-pol and low-temperature experimental data indicates that the student accurately captures the quantum nuclear fluctuations of O--H bonds in water.
Bead-convergence tests and the temperature-resolved $\langle E_K \rangle$ values across 240--370~K are reported in SI Tables~S6 and S7. \\

\noindent This system shows that distillation can transfer the accuracy of an expensive composite teacher (here, a baseline+$\Delta$ MACE) into a single compact student (ACE, ${\sim}10\times$ faster) without loss of thermodynamic, structural, transport, or quantum-nuclear accuracy across a temperature range substantially broader than the one for which teacher model can routinely access via simulation.

\subsection{Proton transfer at the TiO$_2$-water interface}

\begin{figure}[h]
\centering
\includegraphics[width=0.85\textwidth]{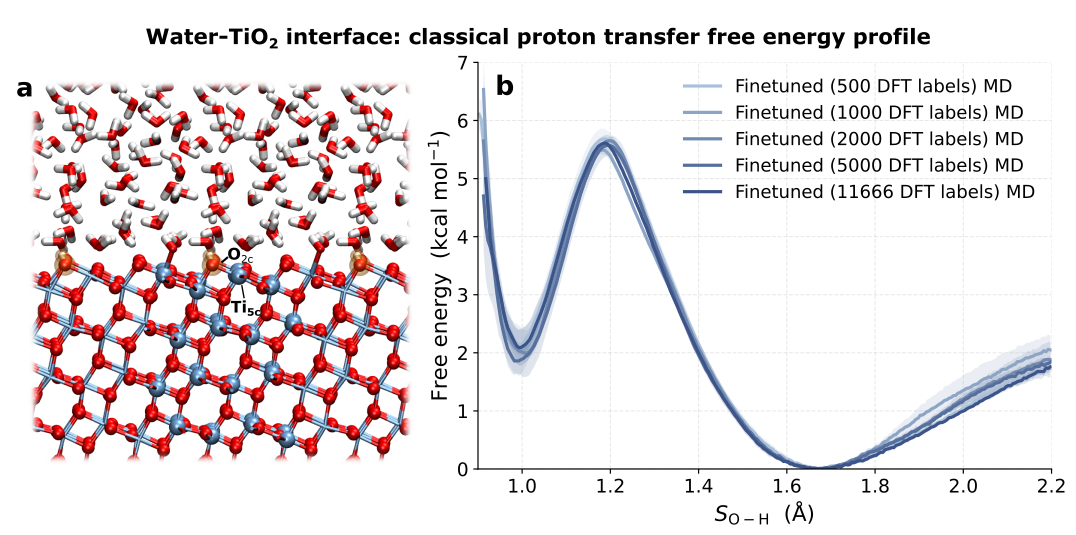}
\caption{(a) Representative atomistic snapshot of the anatase TiO$_2$(101)--water interface used for the umbrella sampling simulations, showing the slab with the adsorbed water layer above. Titanium atoms are shown in blue, oxygen atoms in red, and hydrogen atoms in white. (b) Free energy profiles for water dissociation on the pristine anatase TiO$_2$(101)--water interface from umbrella sampling, using large MACE foundation models finetuned on different numbers of DFT configurations drawn from the Zeng et al.\ dataset.\cite{zengMechanisticInsightWater2023} The minimum near $S_\mathrm{O-H} \approx 1.67$~\AA{} corresponds to the molecularly adsorbed state, and the minimum near $S_\mathrm{O-H} \approx 1$~\AA{} corresponds to the dissociated state. Curves are coloured from light to dark blue with increasing training-set size. Shaded regions denote statistical uncertainty estimated by block averaging.}
\label{fig:interface_teacher_data_efficiency}
\end{figure}

\noindent Finally, we test the combined finetuning and knowledge-distillation protocol to study the role of quantum nuclear motion on water dissociation at the TiO$_2$ interface.
This problem provides a stringent test case because it requires accurately capturing bond breaking and formation in the presence of quantum proton fluctuations at a chemically diverse interface.
%
%
The TiO$_2$-water interface is an intensively studied model system for photocatalysis,\cite{fujishimaElectrochemicalPhotolysisWater1972,thompsonSurfaceScienceStudies2006,zhangTiO2basedSschemePhotocatalysts2024} photoelectrochemistry,\cite{fujishimaElectrochemicalPhotolysisWater1972,thompsonSurfaceScienceStudies2006} environmental remediation,\cite{zhangTiO2basedSschemePhotocatalysts2024} wetting,\cite{laiRecentAdvancesTiO2Based2016,liuBioInspiredTitaniumDioxide2014} and biomaterials.\cite{jafariBiomedicalApplicationsTiO22020,kumarTiO2ItsComposites2018}
%
%
Over the past few decades, experimental work has progressed from ultrahigh-vacuum studies of well-defined single-crystal surfaces to increasingly realistic investigations of hydrated and fully aqueous interfaces using ambient-pressure and surface-sensitive probes.\cite{hendersonInteractionWaterSolid2002,dieboldSurfaceScienceTitanium2003,kettelerNatureWaterNucleation2007} 
These studies have shown that the interfacial speciation of water on TiO$_2$ is sensitive to surface orientation, defects, hydroxylation, coverage, and the surrounding hydrogen-bond network~\cite{hendersonInteractionWaterSolid2002,dieboldSurfaceScienceTitanium2003}. 
%
State-of-the-art simulations of water dissociation at the TiO$_2$ interface have been performed by ~\citet{zengMechanisticInsightWater2023} utilizing MLIPs with enhanced sampling to estimate the classical free energy surface associated with water dissociation.
Given the important role of quantum nuclear effects in proton-transfer reactions~\cite{ceriotti_nuclear_2013}, and the non-trivial influence of interfaces on the stabilization of protonated and hydroxylated species, we study how quantum nuclear motion modulates the thermodynamics of water dissociation at the TiO$_2$ interface.
Dissociation is investigated using the same collective variable $S_\mathrm{O-H}$ adopted in earlier work on this system,\cite{andradeFreeEnergyProton2020,zengMechanisticInsightWater2023} a soft-min approximation to the minimum distance between a particular surface oxygen atom and any hydrogen atom in the system (full definition in Methods). 
Values of $S_\mathrm{O-H} \approx 1$~\AA{} correspond to the dissociated state, where a proton has bonded directly to this surface oxygen, and $S_\mathrm{O-H} \approx 1.67$~\AA{} corresponds to the molecularly adsorbed state. \\

\noindent Due to the high cost of enhanced-sampling simulations, the substantial overhead of path-integral simulations, and the sensitivity of proton delocalization to the accuracy of the potential energy surface~\cite{wang_quantum_2014}, an approach that combines accuracy with computational efficiency is necessary.
To this end, we first obtain an accurate teacher model.
Rather than using the MLIP of ~\citet{zengMechanisticInsightWater2023} directly as the teacher, which was trained on over 10,000 configurations, we instead exploit the data efficiency and robustness of transfer learning by finetuning the MACE-MP-0 foundation model on subsets of that dataset and reproducing the classical free energy surface of water dissociation.
Energy and force test-set parity for the finetuned models are reported in SI Figs.~S8 and S9.
%
%
In Fig.~S12 of the SI, we also compare the density and orientation distributions with the reference produced by ~\citet{zengMechanisticInsightWater2023} and observe all finetuned models to be in agreement.
In Fig.~\ref{fig:interface_teacher_data_efficiency}, we further estimate the classical free energy profiles for water dissociation obtained with the different training subsets using umbrella sampling simulations with 16 windows and 1\,ns long MD per window (more details in the methods section). 
We observe only a weak dependence of the free energies of the reactant and barrier relative to the product on training-set size, with the model trained on 500 configurations reproducing the results of the model fine-tuned on the full dataset to within a fraction of a kcal/mol. \\

\begin{table}[h]
    \centering
    \caption{Free energy difference between the molecularly adsorbed and dissociated states, $\Delta G$, and the free energy barrier, $G^\star$, for water dissociation on the anatase TiO$_2$(101)--water interface predicted by different models.}
    \begin{tabular}{c|c|c}
    \hline
       Model & $\Delta G$ (kcal/mol)  &  $G^\star$ (kcal/mol)   \\
    \hline
       Zeng \textit{et al.}               & $1.8 \pm 0.3$ & $5.4 \pm 0.4$ \\
       Finetuned (500 DFT labels)                       & $2.1 \pm 0.3$ & $5.7 \pm 0.2$ \\
       Finetuned (1000 DFT labels)                      & $2.0 \pm 0.4$ & $5.5 \pm 0.1$ \\
       Finetuned (2000 DFT labels)                      & $2.0 \pm 0.1$ & $5.7 \pm 0.1$ \\
       Finetuned (5000 DFT labals)                      & $1.9 \pm 0.2$ & $5.6 \pm 0.2$ \\
       Finetuned (11666 DFT labels)                       & $2.1 \pm 0.3$ & $5.6 \pm 0.1$ \\
       Classical student                  & $2.1 \pm 0.2$ & $5.5 \pm 0.1$ \\
       Small MACE (1000 DFT)              & $0.9 \pm 0.4$ & $4.6 \pm 0.1$ \\
       Quantum student                    & $1.3 \pm 0.1$ & $3.5 \pm 0.1$ \\
    \hline
    \end{tabular}
    \label{tab:interface_pmf}
\end{table}

\begin{figure}[t]
\centering
\includegraphics[width=0.95\textwidth]{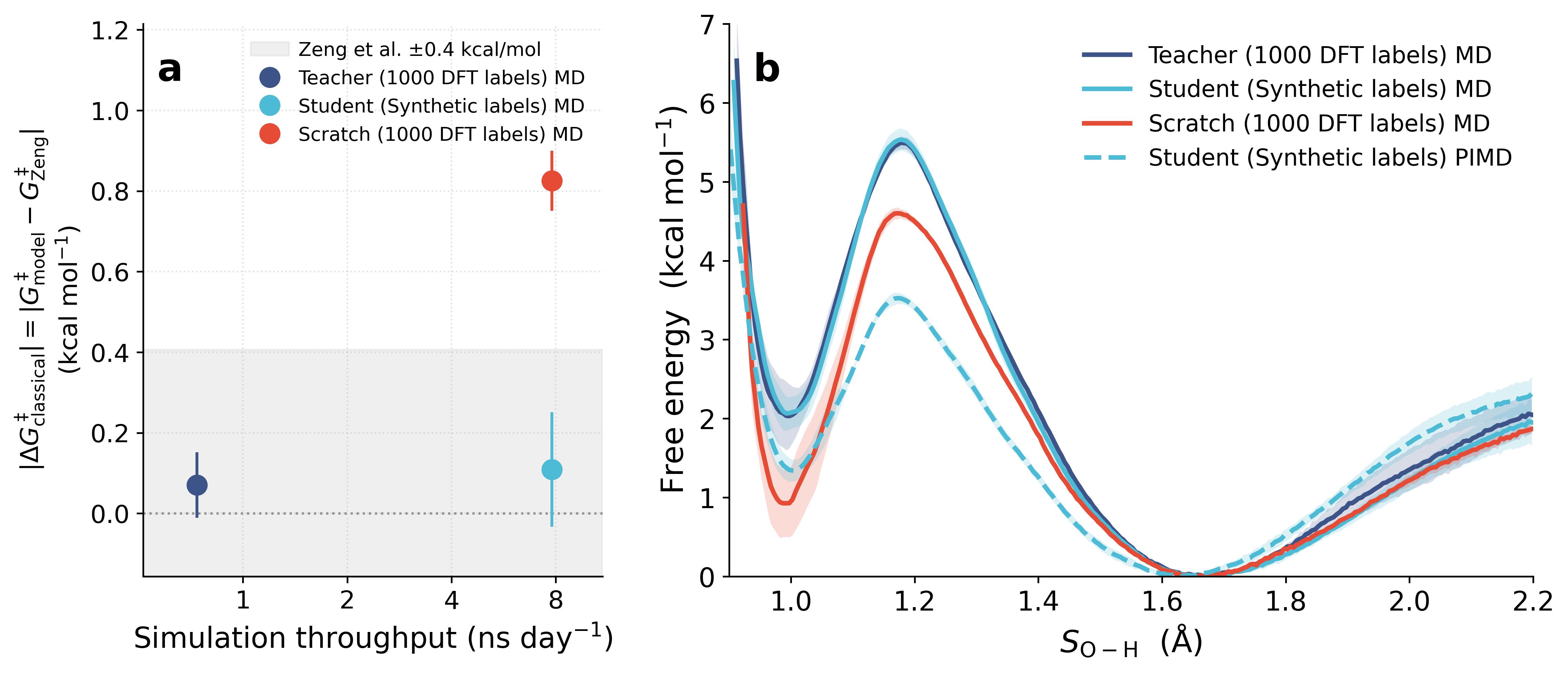}
\caption{Distillation of the classical free energy profile for water dissociation at the TiO$_2$/water interface, and the role of nuclear quantum effects. (a) Speed--accuracy tradeoff for the three classical models: the Finetune-1000 teacher (dark blue circle), the classical distilled student (cyan circle), and a small MACE trained from scratch on the same 1000 DFT configurations (orange-red circle). The y-axis shows the absolute deviation of the dissociation barrier from the value reported by Zeng et al.,\cite{zengMechanisticInsightWater2023} $|\Delta G^\ddagger_\mathrm{classical}| = |G^\ddagger_\mathrm{model} - G^\ddagger_\mathrm{Zeng}|$, with $G^\ddagger_\mathrm{Zeng} = 5.43 \pm 0.41$~kcal/mol. Error bars on each marker show the model's own block-averaging uncertainty in $G^\ddagger$; the gray horizontal band shows the reported uncertainty of $G^\ddagger_\mathrm{Zeng}$. Throughputs are measured on a single NVIDIA A100-PCIe GPU. (b) Free energy profiles along the umbrella sampling coordinate $S_\mathrm{O-H}$ for the same three classical models and the quantum (TRPMD) distilled student (cyan, dashed). Profiles are referenced to the molecularly adsorbed state minimum near $S_\mathrm{O-H} \approx 1.67$~\AA; the dissociated state lies near $S_\mathrm{O-H} \approx 1$~\AA. Shaded bands show statistical uncertainty from block averaging. The quantum student is omitted from (a) because its barrier difference reflects nuclear quantum effects rather than a model-accuracy error.}
\label{fig:interface_distillation_classical_quantum}
\end{figure}

\noindent Having established a data-efficient route to obtain a teacher model, we next ask whether the classical free energy landscape of the model obtained by finetuning MACE-MP-0 on 1000 DFT configurations, can be transferred to a compact student.
To this end, the teacher model is used to label 8000 configurations extracted from classical umbrella-sampling trajectories of the dissociation reaction, collected from 16 umbrella windows run for 50 ps each with snapshots stored every 100 fs, and these data are used to train a compact student MACE model with $L=0$ and 64 channels.
As shown in Fig.~\ref{fig:interface_distillation_classical_quantum}, the distilled student reproduces the classical free energy surface of the teacher up to statistical uncertainty with nearly a factor of 10 speed-up. 
As a control, we also consider a directly trained small model on 1000 configurations (same as used to develop the finetuned teacher model) and note that that while it results in stable enhanced sampling simulations, it yields a smaller barrier and overstabilizes the dissociated form of water.
These results show that transfer learning followed by distillation can transfer the relevant physical knowledge from a high-capacity teacher to a compact student, enabling accurate, robust, and computationally efficient reactive simulations in cases where an equally compact model trained directly on the available reference data would be insufficient. \\

\noindent We next investigate whether the same strategy can be extended to quantum nuclei and whether NQEs modify the free energy profile of water dissociation at the interface.
As a first step, we test whether the classical distilled student can be used directly in path-integral molecular dynamics.
For the present student architecture, namely MACE with $L=0$ and 64 channels, this attempt fails, with the trajectory becoming unstable after only a few femtoseconds.
This shows that, at this model capacity, a student trained only on classical configurations is not transferable to the broader quantum nuclear configuration space explored during interfacial proton transfer.
We therefore construct a separate quantum training set using the teacher by collecting 8000 teacher-labeled configurations from 50 ps PIMD umbrella trajectories across the dissociation coordinate.
We then train a new student with the same architecture as the classical student, namely MACE with $L=0$ and 64 channels, on these quantum configurations.
This model also remains unstable in PIMD, showing that, for this student architecture, access to short teacher-labeled quantum trajectories alone is still insufficient.
To broaden the quantum training domain, we next randomly select 1600 configurations from the PIMD trajectories and generate 6400 additional structures by rattling them with the \textit{augment-atoms} code,\cite{gardnerJlagardnerAugmentatoms2026,gardnerDistillationAtomisticFoundation2025a} again yielding 8000 teacher-labeled quantum configurations in total.
We first retrain the $L=0$, 64-channel student on this augmented dataset, but it again remains unstable in PIMD.
This indicates that the difficulty does not arise solely from limitations of the training set, and that the compact student itself is not sufficiently expressive for the quantum problem.
We therefore next consider a slightly more expressive student, namely MACE with $L=1$ and 32 channels, which is only marginally more expensive than the $L=0$, 64-channel model. A single-point force evaluation timed with ASE on a representative snapshot takes 0.111~s for the $L=1$, 32-channel model versus 0.108~s for the $L=0$, 64-channel one (${\sim}3\%$ slower).
This model produces stable umbrella-sampling simulations across all windows and is therefore adopted as the final quantum student.
Although a full teacher-level PIMD free energy profile is prohibitively expensive, 50 ps unbiased PIMD simulations started from the same initial configuration and velocities give water density profiles and orientational profiles that agree within the noise of the distributions (SI Fig.~S17).
%
%
Together, these results show that extending distillation to NQEs requires both a broader training coverage of the relevant configuration space and a sufficient student expressivity to represent it. \\

\noindent As shown in Fig.~\ref{fig:interface_distillation_classical_quantum}, the quantum free energy profile exhibits both a lower barrier for the molecular-to-dissociative transformation and a smaller free energy difference between the molecularly adsorbed and dissociated states than the classical profile.
NQEs therefore not only soften the barrier region but also stabilize the dissociated state relative to the molecularly adsorbed one.
Both the transition state and the dissociated and partially-dissociated states involve short O$\cdots$H configurations in which the proton is shared between a water oxygen and a surface oxygen; quantum delocalization of the proton over these proton-shared bonds redistributes statistical weight from the localized molecular minimum into the bridged transition-state region and into partially-dissociated configurations, simultaneously lowering the effective barrier and stabilizing the dissociated side.
This delocalization-driven softening of short hydrogen bonds is a generic feature of NQEs in condensed-phase hydrogen-bonded systems.\cite{ceriottiNuclearQuantumEffects2016c,morroneNuclearQuantumEffects2008,marklandNuclearQuantumEffects2018}
Related NQE-driven barrier softening for water dissociation has been reported on stepped Pt(221), Pt(111), and Ru(0001).\cite{litmanDecisiveRoleNuclear2017,lanIonizationWaterQuantum2020,caoQuantumDelocalizationEnables2025} \\

\noindent The computed thermodynamics can be compared with the solid-state $^{17}$O NMR measurements of Yang \textit{et al.},\cite{yangUnravelingAtomicStructure2024} who quantified surface oxygen speciation on anatase TiO$_2$ nanoparticles ($\approx 95\%$ (101)-exposed, $\sim 20$~nm) at ambient temperature ($\approx 298$~K) as a function of H$_2^{17}$O loading.
At $3.1$--$4.7$~mmol~g$^{-1}$, the populations of non-protonated bridging O$_{2c}$ and total surface hydroxyl reach a loading-independent plateau, corresponding to Ti$_{5c}$ saturation and additional physisorbed water layers.
This high-loading regime is the closest experimental analogue of the flat-slab, bulk-water interface simulated here.
Using Yang \textit{et al.}'s coverage-loading calibration ($\approx 28\%$ Ti$_{5c}$ coverage at $0.3$~mmol~g$^{-1}$ and $\approx 46\%$ at $0.5$~mmol~g$^{-1}$), this regime lies well above Ti$_{5c}$ saturation; each Ti$_{5c}$ site is therefore occupied by either chemisorbed or dissociated water, and the bridging-OH fraction relative to total surface O$_{2c}$ gives the dissociated fraction $P_{\mathrm{diss}}$.
Because the ``OH'' curve in Fig.~3c of that work includes both bridging and terminal hydroxyl $^{17}$O resonances, and each dissociation event produces one of each, we take $P_{\mathrm{diss}} = \tfrac{1}{2}\mathrm{OH}$.
Digitizing Fig.~3c at $4.7$~mmol~g$^{-1}$ gives $\mathrm{OH} \approx 0.19$ and $\mathrm{O}_{2c} \approx 0.90$, with $\mathrm{O}_{2c} + \tfrac{1}{2}\mathrm{OH} \approx 1.00$, corresponding to $P_{\mathrm{diss}} \approx 10\%$.
Using $\Delta G = -k_{\mathrm B} T \ln(P_{\mathrm{diss}}/P_{\mathrm{mol}})$ gives $\Delta G \approx +1.3$~kcal~mol$^{-1}$ at $298$~K.
Our free energy profile is computed at $330$~K to match the classical reference of Zeng \textit{et al.};\cite{zengMechanisticInsightWater2023} we assume that this $30$~K offset from the experimental temperature has only a minor effect on the equilibrium $\Delta G$ relative to the other uncertainties in comparing an ensemble-averaged NMR population with a free energy profile computed for an ideal hydrated slab.
Within these uncertainties, the NMR-derived value agrees with the PIMD prediction of $\approx +1.3$~kcal~mol$^{-1}$ and is substantially below the classical prediction of $\approx +2.1$~kcal~mol$^{-1}$ (Fig.~\ref{fig:interface_distillation_classical_quantum}, Table~\ref{tab:interface_pmf}). \\

\noindent A complementary kinetic isotope signature has been reported by Tan \textit{et al.},\cite{tanInterfacialHydrogenBondingDynamics2020a} who used low-temperature scanning tunneling microscopy to follow the dehydrogenation of single submonolayer water molecules on rutile TiO$_2$(110) at $80$--$100$~K, and observed H$_2$O dehydrogenation rates $2$--$4$ times faster than D$_2$O.
They attributed this kinetic isotope effect to the larger zero-point energy of the O--H stretching mode reducing the effective activation barrier.
Although the surface termination, water environment, temperature, and the kinetic versus thermodynamic nature of the observable preclude direct comparison with our equilibrium $G^\star$, the sign and ZPE-based mechanism of the isotope effect are consistent with the NQE-induced barrier reduction observed in our quantum free energy profile.\cite{marxProtonTransfer2002006g} \\

%
\noindent A key insight from this system is that successful distillation is governed by both training-data coverage and student expressivity, and that these two factors must scale in tandem as the target observable becomes more challenging.
For the classical reactive case, the Finetune-1000 teacher transfers cleanly to a compact $L=0$ student trained on configurations drawn from short teacher-driven MD trajectories across all umbrella-sampling windows.
For the quantum problem, the analogous PIMD-trajectory dataset is no longer sufficient on its own: an $L=0$ student augmented with  rattled PIMD configurations is unstable, and an $L=1$ student trained on PIMD-only configurations without rattling is likewise unstable; only the combination of augmented configurational coverage and $L=1$ messages yields a PIMD-stable student, which then reveals an NQE-induced reduction of the dissociation barrier of ${\sim}2$~kcal/mol relative to the classical profile.

\section{Conclusions}

\noindent In this work, we combined transfer learning and knowledge distillation to construct compact MLIPs for three condensed-phase water problems that move from an equilibrium bulk observable to quantum liquid properties and a reactive free energy profile at an interface.
For ice Ih, a student distilled from a fine-tuned MACE teacher is trained only on configurations derived from the low-temperature dataset, but still reproduces the teacher $NPT$ density at a higher temperature while being approximately an order of magnitude faster.
Models of the same size trained directly on the limited DFT dataset either do not sustain stable MD under these conditions or give larger density errors.
For liquid water, a compact ACE student distilled from a $\Delta$-learned CCSD(T)-quality teacher is also approximately an order of magnitude faster than the teacher, yet reproduces the density isobar, RDFs, diffusion coefficient, and nuclear quantum kinetic energies over 240--370~K.
The student captures both the subtle effect of NQEs on the density isobar and the much larger quantum contribution to the hydrogen and oxygen kinetic energies, while retaining agreement with experiment where direct comparisons are available and with MB-pol PIMD benchmarks for the nuclear kinetic energies.
For water dissociation at the anatase TiO$_2$(101)--water interface, distillation transfers the classical umbrella-sampling free energy profile from a fine-tuned teacher to a compact student and makes PIMD umbrella sampling practical for incorporating NQEs into the reaction thermodynamics.
The resulting quantum free energy profile shows that NQEs change both the barrier and the relative stability of the molecular and dissociated states, bringing the computed reaction thermodynamics into closer agreement with experiment. \\

\noindent The three systems show that the requirements for distillation are not fixed, but depend on the observable and on the configuration space sampled during production.
For the classical TiO$_2$/water free energy profile, a compact $L=0$, 64-channel MACE student trained on teacher-labeled umbrella-sampling configurations reproduces the teacher within statistical uncertainty.
To include NQEs at the interface, the same student architecture is not sufficient.
The $L=0$ student remains unstable even when trained on PIMD configurations after additional structures were generated by rattling, whereas an $L=1$, 32-channel student trained on the augmented quantum dataset supports stable PIMD umbrella sampling.
This provides a concrete example in which both the training set and the student architecture must be changed when moving from classical reactive sampling to quantum nuclear sampling. \\

\noindent The TiO$_2$/water application also gives a chemical result that would be substantially more expensive to obtain directly with the larger teacher model.
The quantum student predicts that NQEs lower the water-dissociation barrier by approximately 2~kcal/mol relative to the classical profile and reduce the molecular--dissociated free energy difference from $\approx +2.1$ to $\approx +1.3$~kcal~mol$^{-1}$.
This quantum value matches the $\approx +1.3$~kcal~mol$^{-1}$ estimate inferred from recent solid-state $^{17}$O NMR measurements of water dissociation populations on hydrated anatase TiO$_2$ nanoparticles, whereas the classical value is higher by approximately 0.8~kcal~mol$^{-1}$.
This agreement is obtained despite the uncertainties associated with comparing an ideal flat-slab simulation at 330~K to nanoparticle measurements near ambient temperature, and is consistent with nuclear quantum motion stabilizing dissociated or partially dissociated interfacial configurations in this system. \\

\noindent Taken together, these results show that knowledge distillation can make fine-tuned DFT-quality and $\Delta$-learned CCSD(T)-quality MLIPs usable in simulations that require demanding sampling methods, including long constant-pressure trajectories, PIMD, and umbrella sampling.
We think this opens the door to many problems where accurate potentials must be combined with extensive sampling, especially in catalysis.
Several directions naturally follow from these results.
First, the present workflow constructs the distillation dataset in a one-shot manner and then tests whether the student remains stable in the target simulation.
Coupling distillation with active learning or uncertainty-guided sampling would make this process more systematic, allowing configurations visited by the student during production to trigger additional teacher labeling when needed.
Second, the interface calculations show that test-set errors alone do not determine whether a student will support stable PIMD or enhanced-sampling simulations.
A more quantitative connection between the complexity of the sampled configuration space and the minimum required student architecture would make model selection less empirical.
Third, the TiO$_2$/water teacher was fine-tuned on classical-nuclei DFT configurations, while the quantum student was trained to cover the additional configurations sampled by PIMD.
This strategy is sufficient for the present system, but for reactions where NQEs open qualitatively different pathways, the teacher itself may need to be fine-tuned or validated on configurations sampled with quantum nuclei.
Finally, applying the same approach to other reactive interfaces, proton-coupled processes, solution reactions, and ion-transport problems will test whether the coverage and architecture trends observed here are general across condensed-phase chemistry.

\section{Methods}\label{sec:methods}

\noindent This work combines transfer learning and knowledge distillation for three representative condensed phase problems, presented in order of increasing difficulty: a solid (ice Ih), a liquid (water), and a reactive solid--liquid interface (water dissociation at the anatase TiO$_2$(101)--water interface). In all three cases, we first construct or adopt an accurate teacher model for the target chemical domain and then train smaller student models on teacher-labeled configurations. \\

\subsection{Reference data and target systems}

\noindent For the solid system, we use the ice Ih dataset of Kaur \textit{et al.}, consisting of 400 revPBE-D3(0)\cite{zhangCommentGeneralizedGradient1998,grimmeConsistentAccurateInitio2010}-labeled configurations at 100 K and 1 bar.\cite{kaurDataefficientFinetuningFoundational2025} These data are used both to finetune a large MACE foundation model and to construct DFT-only control models. \\

\noindent For the liquid system, the teacher is the baseline+$\Delta$ MACE model of O'Neill \textit{et al.}\cite{oneillRoutineCondensedPhase2025,FastgroupcamData_CC_water2026}, which was designed to reproduce CCSD(T)-quality condensed phase water behavior through a combination of periodic DFT data and gas phase CCSD(T)$-$DFT energy differences. The corresponding student model is trained on configurations labeled by this teacher. \\

\noindent For the interface system, we use the optB88-vdW\cite{klimesChemicalAccuracyVan2009} dataset of Zeng \textit{et al.}\cite{zengMechanisticInsightWater2023,chengBingqingChengTiO2water2026} We retain only the TiO$_2$--water interface configurations involving pristine and defective anatase and rutile surfaces, since these are the environments relevant to the present TiO$_2$--water problem. The resulting dataset contains 12\,963 configurations. As described in the Results and Discussion section, the final benchmark is performed exclusively on the pristine anatase TiO$_2$(101)--water interface, while finetuning is carried out on randomly selected subsets drawn from the broader interfacial dataset. \\

\subsection{Teacher models and student models}

\noindent All finetuning and student-training procedures use a random 9:1 training:validation split.
For finetuning in the solid (ice Ih) and reactive interface (TiO$_2$--water) systems, we start from the MACE-MP-0a(L) foundation model.\cite{batatiaFoundationModelAtomistic2025a,ACEsuitMacefoundations2026} Finetuning is carried out with the MACE package \cite{Batatia2022mace,batatiaDesignSpaceE3equivariant2025,ACEsuitMace2026a} version 0.3.14 using CUDA 12.8. 
For ice Ih, the MACE-MP-0a(L) model is finetuned on the 400 DFT-labeled configurations from Kaur \textit{et al.} This finetuned model is used as the teacher for the ice subsection.
For the TiO$_2$/water interface, 10\% of the available interface configurations are reserved as a strict test set and excluded from all finetuning and student model training. From the remaining configurations, random subsets of size 500, 1000, 2000, 5000, and all available data are selected to finetune separate large MACE teacher models respectively. 
For all subset training runs, a force weight of 10 and an energy weight of 10 are used, together with the default learning rate of 0.01. Finetuning is run for 1000 epochs, with stage two starting at epoch 750 and using a reduced learning rate of 0.00025. 
For when finetuning on all available data Finetune, the same hyperparameters are used except that the total number of epochs is reduced to 500 and stage two starts at epoch 250 because of the larger amount of training data. \\

\noindent Teacher-labeled training configurations are generated in two ways: (i) MD simulations driven by the teacher model, and (ii) rattling existing configurations followed by teacher labeling. \\

\noindent Two model families are used as students: MACE and ACE. For the MACE students, we use two interaction layers, 4-body messages in each layer (correlation order 3), a radial cutoff of 5.0~\AA, and radial features constructed from 10 Bessel functions multiplied by a smooth polynomial cutoff function. The spherical expansion uses $l_{\mathrm{max}}=3$, while $L$ denotes the maximum angular frequency order of the message in the irreducible representation basis.\cite{Batatia2022mace,batatiaFoundationModelAtomistic2025a} For the solid system, two small MACE variants are considered, both with $L=0$ (invariant messages) and channel sizes of 64 and 128, respectively. The two models comprise 148944 and 460432 parameters, respectively. For the interface system, the classical student uses $L=0$ with 64 channels, while the final stable quantum student uses $L=1$ (equivariant messages) with 32 channels and comprises 93776 parameters. Training is run for 600 epochs, with stage two starting at epoch 400. \\

\noindent ACE\cite{drautzAtomicClusterExpansion2019a} students are trained with the \texttt{Pacemaker} package\cite{ICAMSPythonace2026,bochkarevEfficientParametrizationAtomic2022a,lysogorskiyPerformantImplementationAtomic2021} and are used only for the solid and liquid systems. The ACE model uses a radial cutoff of 6.0~\AA, a Finnis--Sinclair-type nonlinear embedding function with 2 atomic properties for the solid system and 6 atomic properties for the liquid system. The number of basis functions is truncated at 1000 per element for the final potential with body orders up to 6. The solid and liquid ACE models comprise 5296 and 14320 parameters, respectively. Hierarchical fitting, which adds parameters iteratively in small batches, is used with a batch size of 100 and $\kappa = 0.9$, which determines the relative contribution of the squared deviation in predicted energies. \\

\noindent For the ice Ih system, 1600 additional structures are generated by rattling the 400 DFT-labeled configurations using the \texttt{augment-atoms} code\cite{gardnerJlagardnerAugmentatoms2026,gardnerDistillationAtomisticFoundation2025a}. These 1600 structures, together with the original 400 configurations, yield a 2000 configuration teacher-labeled dataset used to train the student models. As controls, corresponding small MACE and ACE models are also trained directly on the 400 DFT configurations. \\

\noindent For the liquid water system, the student ACE model is trained on 10\,000 configurations selected from teacher-driven equilibrium trajectories spanning a broad range of thermodynamic conditions, including temperatures up to 360 K to improve coverage of higher-energy configurations. Specifically, training configurations are drawn from 0.5 ns NVT simulations at 260, 300, 330, and 360 K and densities of 0.85, 0.90, 0.95, 1.00, 1.05, 1.10, and 1.15~$\mathrm{g/cm^3}$; 0.5 ns NPT simulations at the same temperatures and pressures of $-1500$, $-500$, $-300$, 1, 2, 500, 1000, 4000, and 8000 bar; and 50 ps thermostatted ring polymer MD (TRPMD) NVT simulations at 300 K over a density scan of 0.85, 0.90, 0.95, 1.00, 1.05, and 1.10~$\mathrm{g/cm^3}$. Some NPT simulations with negative pressures and high temperatures are not stable and therefore abandoned. \\

\noindent For the TiO$_2$/water interface, the classical distillation dataset is generated from short MD trajectories driven by the Finetune-1000 teacher. Specifically, configurations are collected from all umbrella sampling windows, each run for 50 ps, with snapshots stored every 100 fs, giving 8000 teacher-labeled configurations in total. These configurations are used to train the classical student MACE model. As a control, an identically sized small MACE is trained directly on the same 1000 DFT configurations used to finetune the teacher. \\

\noindent For the quantum interface problem, we first collect 8000 teacher-labeled configurations from TRPMD umbrella trajectories of 50 ps per window using 32 beads. A student MACE with the same architecture as the classical student ($L=0$, 64 channels) and a slightly larger variant ($L=1$, 64 channels) are both found to be unstable in subsequent TRPMD simulations. We therefore broaden the quantum training set by randomly selecting 1600 configurations from the TRPMD trajectories and generating 6400 additional structures by rattling them with the \texttt{augment-atoms} code\cite{gardnerJlagardnerAugmentatoms2026,gardnerDistillationAtomisticFoundation2025a}, yielding 8000 teacher-labeled quantum configurations in total. Two additional quantum students are then trained: one with $L=0$, 64 channels, and a second with $L=1$, 32 channels. The latter is the only model that supports stable TRPMD umbrella sampling simulations and is therefore adopted as the final quantum student. \\

\subsection{Molecular dynamics simulations}

\noindent Classical MD simulations with the finetuned large MACE-MP-0 teacher models are performed with LAMMPS\cite{thompsonLAMMPSFlexibleSimulation2022j} using the ML-IAP package and the \texttt{mliap} pair style. Small and medium MACE student models are interfaced to LAMMPS through the Symmetrix library\cite{wcwittWcwittSymmetrix2026,kovacsMACEOFFShortRangeTransferable2025b,batatiaFoundationModelAtomistic2025a} using the \texttt{symmetrix/mace} pair style. ACE simulations are carried out in LAMMPS using the ML-PACE package and the \texttt{pace} pair style. TRPMD simulations are performed with i-PI\cite{litmanIPI30Flexible2024,IpiIpi2026} interfaced to LAMMPS. \\

\noindent Unless otherwise stated, classical MD simulations use a timestep of 0.5 fs, and TRPMD simulations use a timestep of 0.25 fs. Exceptions for unstable models are reported in the Results and Discussion section. For classical MD simulations in LAMMPS, a Nos\'e--Hoover thermostat\cite{noseUnifiedFormulationConstant1984c,hooverCanonicalDynamicsEquilibrium1985c} and barostat are used, with damping times of $100 \times \mathrm{d}t$ and $1000 \times \mathrm{d}t$, respectively, where $\mathrm{d}t$ is the simulation timestep. For TRPMD simulations, the \texttt{pile\_g} thermostat\cite{craigQuantumStatisticsClassical2004,rossiHowRemoveSpurious2014a} with $\tau=100~\mathrm{fs}$ and $\lambda=0.5$, together with an isotropic Bussi--Zykova-Parrinello barostat\cite{bussiIsothermalisobaricMolecularDynamics2009} with $\tau=1000~\mathrm{fs}$ for NPT simulations, is used. \\

\noindent For ice Ih, classical NPT simulations are carried out at 100 K and 1 bar and at 220 K and 1 bar. For each model, 5 independent replicas of 1 ns are run after 0.5 ns of equilibration. \\

\noindent For liquid water, classical NPT density calculations are performed from 240 to 370 K in 10 K increments, except that 270 K is replaced by 273.15 K. At each temperature, 5 independent replicas are run until the standard deviation of the density falls below 0.001~$\mathrm{g/cm^3}$, with a maximum trajectory length ranging from 2 ns at high temperature to 20 ns for supercooled water. Quantum density calculations are performed with TRPMD in the NPT ensemble using 32 beads and the same convergence criterion, subject to a maximum trajectory length of 2 ns. For radial distribution function (RDF) and diffusion calculations, classical NVT and TRPMD NVT simulations are performed at the equilibrium densities predicted by the corresponding classical NPT and TRPMD NPT simulations, respectively. Classical NVT calculations use 5 independent 4 ns replicas after 500 ps of equilibration, while TRPMD calculations use 5 independent 0.4 ns replicas after 10 ps of equilibration. For NKE calculations, TRPMD simulations with 64 beads in NVT ensemble are performed, with all other settings identical to the 32 beads case. \\

\noindent For the TiO$_2$--water interface, all MD simulations are performed in the NVT ensemble at 330 K using the same simulation cell as in the anatase(101)--water setup of Zeng \textit{et al.}, in order to ensure direct comparability. The simulation volume is fixed such that the water density at the center of the slab is maintained at 1~$\mathrm{g/mL}$. Short unbiased classical MD trajectories of 1 ns are used to analyze interfacial water density profiles and orientational distributions for the untuned and finetuned large models. Short unbiased TRPMD simulations of 50 ps are used to compare the final quantum student with the Finetune-1000 teacher. The TRPMD NVT settings are the same as those used for liquid water. \\

\subsection{Umbrella sampling for the TiO$_2$--water interface}

\noindent The free energy profile for water dissociation on the anatase TiO$_2$(101)--water interface is computed by umbrella sampling along the same collective variable used in earlier work on this system.\cite{andradeFreeEnergyProton2020,zengMechanisticInsightWater2023} The collective variable $S_{\mathrm{O-H}}$ is a soft-min approximation to the minimum distance between a particular surface oxygen atom and any hydrogen atom in the system:
\[
S_{\mathrm{O-H}} = \frac{\lambda}{\ln \sum_i \exp{(\lambda / r_{i\mathrm{O}})}}, \qquad \lambda = 500~\mathrm{\AA},
\]
\noindent where the index $i$ runs over all hydrogen atoms, $r_{i\mathrm{O}}$ is the distance from the $i$-th hydrogen to the chosen surface oxygen, and the large $\lambda$ ensures that $S_{\mathrm{O-H}}$ is dominated by the smallest $r_{i\mathrm{O}}$. Values of $S_{\mathrm{O-H}} \approx 1$~\AA{} correspond to the dissociated state, where a proton has bonded directly to the surface oxygen, while $S_{\mathrm{O-H}} \approx 1.67$~\AA{} corresponds to the molecularly adsorbed state, where the closest hydrogen belongs to the water molecule sitting above the surface oxygen. \\

\noindent The umbrella centers span 0.96 to 2.10~\AA, and the force constants range from 125 to 800 kcal/mol. Each window is equilibrated for 10 ps before a production run of 1 ns for the teacher model and quantum student, and 2 ns for the classical student. TRPMD umbrella sampling uses 32 beads. Free energy profiles are estimated using WHAM\cite{kumarWeightedHistogramAnalysis1992a,grossfieldWHAMGrossfieldLab}. Statistical uncertainties are estimated by block averaging: each trajectory is divided into three equal parts, three free energy profiles are reconstructed independently, and the standard deviation among them is used as the uncertainty estimate. \\

\section*{Acknowledgements}
We thank all members of the Initiative of Computational Catalysis for valuable comments on this work.
We thank William Witt for making us aware of the computationally efficient symmetric implementation of small MACE models, which substantially improved the performance of some of the student models considered in this work.
VK acknowledges the hospitality of the Initiative for Computational Catalysis at the Flatiron Institute during a visit, where a portion of this research was carried out.
The Flatiron Institute is a division of the Simons Foundation. 

\section*{Conflict of Interest}
The authors declare no conflict of interest. \\

\section*{Data availability}
The training datasets, trained models, and simulation input files used in this work are available at \url{https://github.com/Alexandrina-Chen/data-for-mlip-kd}.

\newpage


\bibliography{reference}

\end{document}